\documentclass[onecolumn,notitlepage,secnumarabic,amssymb, nobibnotes, aps]{revtex4-1}
\usepackage{graphicx}
\usepackage{multirow,amsmath,amssymb,graphicx,epstopdf}

\newcommand{\kB}{k_{FB}}

\newcommand{\cH}{\underline{H}}
\newcommand{\cvH}{\vec{\cH}}
\newcommand{\cvHl}{\cvH_{\ell}}

\newcommand{\cE}{\underline{E}}

\newcommand{\cvE}{\vec{\cE}}

\newcommand{\cS}{\underline{S}}
\newcommand{\cvS}{\vec{\cS}}

\newcommand{\cHl}{\underline{H}_{\ell}}

\newcommand{\lo}{\lambda_{0}}
\newcommand{\nm}{\, \rm{nm}}
\newcommand{\um}{\, \rm{\mu m}}
\newcommand{\Ho}{H_{0}}

\newcommand{\kxo}{k_{x,0}}
\newcommand{\kzo}{k_{z,0}}
\newcommand{\ckzl}{\underline{k}_{z,\ell}}
\newcommand{\ko}{k_{0}}
\newcommand{\eo}{\epsilon_{0}}
\newcommand{\zo}{z_{0}}
\newcommand{\zl}{z_{\ell}}
\newcommand{\dl}{d_{\ell}}
\newcommand{\zcl}{z_{c,\ell}}
\newcommand{\zcmj}{z_{c,mJ+j}}

\newcommand{\yh}{\hat{y}}
\newcommand{\xh}{\hat{x}}
\newcommand{\zh}{\hat{z}}
\newcommand{\ce}{\underline{\epsilon}}
\newcommand{\cm}{\underline{\mu}}
\newcommand{\cel}{\ce_{\ell}}
\newcommand{\cml}{\cm_{\ell}}
\newcommand{\cnl}{\underline{n}_{\ell}}

\newcommand{\cej}{\ce_{j}}
\newcommand{\cmj}{\cm_{j}}
\newcommand{\cnj}{\underline{n}_{j}}

\newcommand{\ckzj}{\underline{k}_{z,j}}
\newcommand{\zjmo}{z_{j-1}}

\newcommand{\cAl}{\underline{A}_{\ell}}
\newcommand{\cBl}{\underline{B}_{\ell}}
\newcommand{\cAmj}{\underline{A}_{mJ+j}}
\newcommand{\cBmj}{\underline{B}_{mJ+j}}
\newcommand{\cAj}{\underline{A}_{j}}
\newcommand{\cBj}{\underline{B}_{j}}
\newcommand{\cUj}{\underline{\textbf{U}}_{j}}
\newcommand{\cAlpo}{\underline{A}_{\ell+1}}
\newcommand{\cBlpo}{\underline{B}_{\ell+1}}
\newcommand{\cAlpN}{\underline{A}_{\ell+s}}
\newcommand{\cBlpN}{\underline{B}_{\ell+s}}

\newcommand{\cWlNl}{\underline{\textbf{W}}_{\ell+s,\ell}}
\newcommand{\cWjJj}{\underline{\textbf{W}}_{j+J,j}}

\newcommand{\cWjo}{\underline{\textbf{W}}_{j,1}}
\newcommand{\cUo}{\underline{\textbf{U}}_{1}}
\newcommand{\cQ}{\underline{\textbf{Q}}}
\newcommand{\cl}{\underline{\boldsymbol{\lambda}}}
\newcommand{\clone}{\underline{\lambda}^+}
\newcommand{\cltwo}{\underline{\lambda}^-}
\newcommand{\clpm}{\underline{\lambda}^\pm}
\newcommand{\cTl}{\underline{\textbf{T}}_{\ell}}
\newcommand{\cPl}{\underline{\textbf{P}}_{\ell}}
\newcommand{\cTq}{\underline{\textbf{T}}_{q}}
\newcommand{\cPq}{\underline{\textbf{P}}_{q}}
\newcommand{\cpl}{\underline{p}_{\ell}}
\newcommand{\celpo}{\ce_{\ell+1}}
\newcommand{\ckzlpo}{\underline{k}_{z,\ell+1}}
\newcommand{\cA}{\underline{A}}
\newcommand{\cB}{\underline{B}}
\newcommand{\cT}{\underline{\textbf{T}}}
\newcommand{\cref}{\underline{r}}
\newcommand{\ctran}{\underline{t}}

\newcommand{\zMJ}{z_{MJ}}
\newcommand{\ctH}{\cH}
\newcommand{\ctEx}{\cE_{x}}
\newcommand{\ctEz}{\cE_{z}}
\newcommand{\zlmo}{z_{\ell-1}}
\newcommand{\zmjmo}{z_{mJ+j-1}}
\newcommand{\ckB}{\underline{k}_{FB}}
\newcommand{\cnB}{\underline{n}_{FB}}

\newcommand{\kapz}{\kappa_z}
\newcommand{\kapx}{\kappa_x}
\newcommand{\cLj}{{\underline{\textbf{L}}_{j}}}
\newcommand{\cL}{{\underline{L}}}
\newcommand{\cLMat}{{\underline{\textbf{L}}}}
\newcommand{\cLone}{{\underline{L}^+_j}}
\newcommand{\cLtwo}{{\underline{L}^-_j}}

\newcommand{\cFB}{{\underline{\textbf{FB}}}}
\newcommand{\cFBone}{{\underline{FB}^+}}
\newcommand{\cFBtwo}{{\underline{FB}^-}}
\newcommand{\cFBpm}{{\underline{FB}^\pm}}
\newcommand{\cC}{{\underline{\textbf{C}}}}
\newcommand{\cCone}{{\underline{C}^+}}
\newcommand{\cCtwo}{{\underline{C}^-}}
\DeclareMathOperator{\rect}{rect}
\DeclareMathOperator{\sinc}{sinc}
\DeclareMathOperator{\sgn}{sgn}

\begin{document}

\title{Fourier-Domain Electromagnetic Wave Theory for Layered Metamaterials of Finite Extent}

\author{Kenneth~J.~Chau}
\affiliation{School of Engineering, The University of British Columbia, Kelowna, British Columbia, Canada}
\author{Mohammed~H.~Al~Shakhs}
\affiliation{School of Engineering, The University of British Columbia, Kelowna, British Columbia, Canada}
\author{Peter~Ott}
\affiliation{Heilbronn University, Heilbronn, Germany}

    \begin{abstract}
     The Floquet-Bloch theorem allows waves in infinite, lossless periodic media to be expressed as a sum of discrete Floquet-Bloch modes, but its validity is challenged under the realistic constraints of loss and finite extent.  In this work, we mathematically reveal the existence of Floquet-Bloch modes in the electromagnetic fields sustained by lossy, finite periodic layered media using Maxwell's equations alone without invoking the Floquet-Bloch theorem.  Starting with a transfer-matrix representation of the electromagnetic field in a generic layered medium, we apply Fourier transformation and a series of mathematical manipulations to isolate a term explicitly dependent on Floquet-Bloch modes.  Fourier-domain representation of the electromagnetic field can be reduced into a product of the Floquet-Bloch term and two other matrix factors: one governed by reflections from the medium boundaries and another dependent on layer composition.   Electromagnetic fields in any finite, lossy, layered structure can now be interpreted in the Fourier-domain by separable factors dependent on distinct physical features of the structure.  The developed theory enables new methods for analyzing and communicating the electromagnetic properties of layered metamaterials.
    \end{abstract}

\maketitle

\section{Introduction}
Recent contributions to the tradition of inquiry into the electromagnetic properties of planar layered structures~\cite{Brillouin1953,Rytov1956,Ishimaru1965,Berreman1972,Yeh1977a,Yariv1977,Yeh1979,Yeh1988,Macleod2010} have been sparked by the novel conceptualization of these structures -- particularly those composed of sub-wavelength-thick layers of metal -- as metamaterials.  The metamaterial concept is used to describe a structure with sub-wavelength scale heterogeneity in terms of plane-wave parameters such as refractive index and impedance.  Planar layered structures, which possess heterogeneity along just a single direction, are the simplest metamaterial form and provide an experimentally feasible template for metamaterial devices operating at visible frequencies and beyond.  This is largely due to the availability of thin-film deposition techniques with layer thickness control on sub-nanometer scales.

The classification of layered structures as metamaterials begins by seeking an analog homogeneous structure with a plane-wave solution that mimics the more intricate wave solution corresponding to the original structure (this process is known as homogenization).  Each homogenization technique invokes a set of assumptions, which are not always justified, to arrive at its plane-wave parameters. Effective medium theory can be used to define an effective permittivity tensor through volumetric averaging of the local permittivity values~\cite{Rytov1956,Sihvola1999,Born1970}, which, for a layered structure, simplifies to a thickness-weighted average of the layer permittivity values.  Although effective medium theory is intuitive, it relies upon the electrostatic approximation which neglects time-derivative terms in Maxwell's equations.  The scattering parameter method~\cite{Nicholson1970,Weir1974,Smith2005,Pozar2005} is based on equating the reflection and transmission coefficients of a heterogeneous structure to those of an equivalent homogeneous structure.  Drawbacks include non-uniqueness~\cite{Mortensen2009,Breinbjerg2013} and the absence of correlation to the fields inside the structure.  It is possible to derive effective constitutive parameters by averaging local permittivity values weighted by the fields~\cite{Smith2006} or energy densities~\cite{Chau2012} inside a structure, although this method also suffers from non-uniqueness.

One of the most powerful methods for describing the fields in a periodic structure is based on the Floquet-Bloch theorem~\cite{Floquet1883,Bloch1928}, which enables the wave inside a structure to be decomposed into a set of Floquet-Bloch modes $\kB$~\cite{Ramo1965,Chu1977,Gralak2000,Lombardet2005A,Lombardet2005B,Kishk2007,Mesa2008}.  If one mode carries dominant power, it is assumed to approximate the entire wave and its plane-wave parameters are conferred to the structure~\cite{Sjoberg2006,Tsukerman2008,Cabuz2008,Rockstuhl2008,Alu2011,Andryieuski2012}.  The Floquet-Bloch modes of a periodic layered structure can be found by imposing translational invariance of the wave over a period within a multiplicative exponential factor~\cite{Yeh1988}, a procedure that implicitly assumes infinite extent and no loss.  When the medium is lossy, the Floquet-Bloch modes are complex-valued~\cite{Fan1996,Huang2004,Parisi2012}, but no longer discrete~\cite{Sjoberg2006}.  When the medium is finite, translational invariance is altogether lost for two reasons: reflections from the end facets of the medium and dependence on the excitation conditions in the external bounding media~\cite{Gralak2000}.  Recently, Fourier transformation of the numerically-computed wave solution in a finite-sized, lossless periodic layered medium has revealed a correlation between the spatial-frequency content of the wave and the Floquet-Bloch modes corresponding to an unbounded version of the periodic medium~\cite{Mortensen2009,Clausen2014}.  This correlation, however, has not been made mathematically explicit and has yet to be established for the case of loss, a condition vital for analysis of metamaterials which frequently incorporate metals.

In this work, we mathematically prove using only Maxwell's equations that Floquet-Bloch modes are inherent in the electric and magnetic fields sustained by lossy periodic layered media of finite extent.  Starting with a standard transfer-matrix representation of the electromagnetic field, we apply analytical Fourier transformation and a series of manipulations to identify and isolate Floquet-Bloch modes in the field solution, without invoking the Floquet-Bloch theorem.  Whereas the electromagnetic field in an infinite, lossless periodic medium consists of a discrete sum of modes, the field in a finite, lossy structure consists of a continuum of modes, whose distribution is governed by an interplay between the effects of boundary reflections and layer composition.  We condense the Fourier-domain representation into a compact product of three matrix factors, each dependent on physically distinctive parameters.  The theory developed here provides new insights into the role of Floquet-Bloch modes in lossy, finite periodic media and can help further refine homogenization methods for describing layered metamaterials.

\section{Generalized Description of a Layered Medium}

\begin{figure*}
\begin{center}
\includegraphics[scale=0.80]{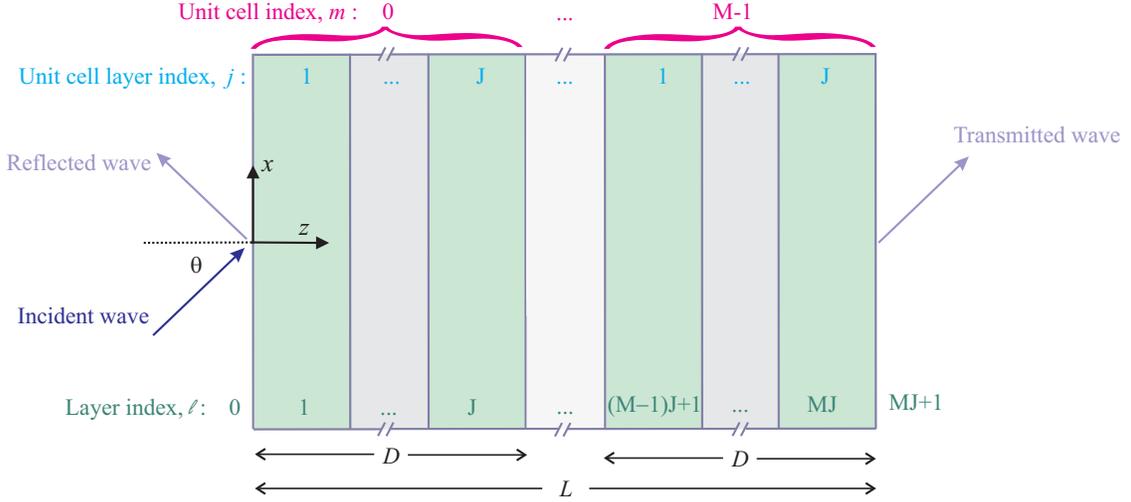}
\caption{\label{Geom} Geometry under consideration consisting of a one-dimensional periodic layered medium bounded by two semi-infinite half spaces and composed of $M$ repeated unit cells, each consisting of $J$ layers. The medium is excited from one half-space by an incident plane inclined at an arbitrary angle $\theta$ in the $xz$ plane. }
\end{center}
\end{figure*}

We consider a generalized one-dimensionally periodic medium (Figure~\ref{Geom}) bounded by two semi-infinite half spaces and composed of $M$ repeated unit cells, each consisting of $J$ layers, yielding a total of $MJ$ layers.  The unit cells are referenced by the integer $m=0,...,M-1$ and the layers within any unit cell are referenced by the integer $j = 1, ... , J$.  Each layer in the medium is uniquely labeled by the integer $\ell = m J + j = 1, ... , MJ$ and, for sequential consistency, the free-space half-spaces to the left and right of the medium are labeled $\ell = 0$ and $\ell = MJ + 1$, respectively.  Layer $\ell$ has a thickness of $\dl$ and its linear electromagnetic properties are generally specified by a complex-valued relative permittivity $\cel$ (the underline denotes a complex variable) and a complex-valued relative permeability $\cml$, resulting in a complex-valued refractive index~\cite{Depine2004}
\begin{equation}\label{handcondition}
\cnl = \sgn(\Re[\cel]|\cml|+\Re[\cml]|\cel|)\sqrt{\cel \cml},
\end{equation}
which is permitted to have a real part that is positive (right-handed) or negative (left-handed).  The total thickness of the layered medium is $L = \sum_{\ell = 1}^{MJ}\dl$.  Due to periodic repetition of the unit cell, the quantities $\dl$, $\cel$, $\cml$, and $\cnl$, corresponding to layer $j$ of an arbitrary unit cell $m$, can be equivalently denoted $d_j$, $\cej$, $\cmj$, and $\cnj$ corresponding to layer $j$ of unit cell $m=0$.  The plane of the layers is aligned parallel to the $xy$ plane and we denote the location of the plane between layer $\ell$ and $\ell + 1$ as $\zl$, setting the position of the plane of the first interface $\zo=0$ without loss of generality.

An electromagnetic plane wave is incident onto the medium from the half-space $z<0$ inclined at an angle $\theta$ in the $xz$ plane.  Due to the independence of this configuration with respect to the $y$ coordinate, any solution can be expressed as a linear combination of solutions obtained by assuming either transverse-electric (TE) polarization (electric field $\cvE$ aligned to the $y$ axis) or transverse-magnetic (TM) polarization (magnetic field $\cvH$ aligned to the $y$ axis).  Here, we treat the case of TM polarization, noting that the transformations $\cvE \rightarrow -\cvH$, $\cvH \rightarrow \cvE$, and $\ce \rightleftarrows \cm$ yield complimentary equations for TE polarization~\cite{Kong2005}.

\section{Transfer-Matrix Representation of the Electromagnetic Field} \label{spatialsoln}

We derive a representation of the electromagnetic wave solution in the finite one-dimensional periodic layered structure for the case in which one of the end facets is subject to plane-wave illumination.  An incident TM-polarized electromagnetic wave is given by $\cvH =  \Ho e^{i (\kxo x + \kzo z)} \yh$, where $\Ho$ is the amplitude, $\kxo$ and $\kzo$ are the real-valued wave-vector components along the $x$- and $z$-axes, respectively, and $\ko = \sqrt{\kxo + \kzo}$ is the wave vector in the left half space labeled ``0''.  The wave is time-harmonic where an $e^{-i \omega t}$ dependence is assumed but suppressed.  Invoking field continuity across the interfaces, the magnetic field in an arbitrary layer $\ell$ can be written as a sum of two counter-propagating waves using matrix formalism
\begin{equation}\label{Hl}
\begin{split}
\cvHl(x,z) &= \cHl(x,z) \yh \\
&=  e^{i \kxo x} \left(
	\begin{array}{c} e^{i \ckzl (z-\zl)} \\
		e^{-i\ckzl (z-\zl)}
	\end{array}\right)^T  \left(
	\begin{array}{c} \cAl \\ \cBl \end{array}\right) \yh,
\end{split}
\end{equation}
where $T$ denotes the transpose operator, $\cAl$ and $\cBl$ are the wave coefficients, and $\ckzl$ is the wave-vector component in layer $\ell$ along the $z$-axis (note that $\ckzl$ corresponding to layer $j$ of an arbitrary unit cell $m$ can be equivalently denoted $\ckzj$ corresponding to layer $j$ of unit cell $m=0$).  The wave-vector component $\ckzl$ is related to the layer refractive index by
\begin{equation}
\ckzl = \cnl \sqrt{\ko^2 - \left(\frac{\kxo}{\cnl}\right)^2 },
\end{equation}
where, according to (\ref{handcondition}), $\cnl$ can have a real part that is either positive or negative, describing a right- or left-handed medium, respectively.

The wave solution can be solved by relating the wave coefficients $\cAl$ and $\cBl$ across the $MJ+1$ boundaries. The wave coefficients in an arbitrary layer $\ell$ can be related to the coefficients in an adjacent layer $\ell+1$ by
\begin{equation}
\left(\begin{array}{c} \cAlpo \\ \cBlpo \end{array}\right) = \cTl \cPl  \left(\begin{array}{c} \cAl \\ \cBl \end{array}\right),
\end{equation}
where the propagation matrix $\cPl$ (bold denotes a matrix quantity) corresponding to layer $\ell$ is given by
\begin{equation}
\cPl = \begin{pmatrix} e^{i\ckzl\dl} & 0\\ 0 & e^{-i\ckzl\dl} \\ \end{pmatrix},
\end{equation}
and the transmission matrix $\cTl$ corresponding to the interface between layer $\ell$ and $\ell+1$ is given by
\begin{equation}
\cTl = \frac{1}{2} \begin{pmatrix} 1+\cpl & 1-\cpl\\ 1-\cpl & 1+\cpl \\ \end{pmatrix},
\end{equation}
with $\cpl = (\celpo \ckzl)/(\cel \ckzlpo)$.  Assuming uni-directional wave excitation from the left half-space, the wave coefficients in layer 1 are related to the coefficients in the left half-space by
\begin{equation} \label{coeff_one}
\left(\begin{array}{c} \cA_{1} \\ \cB_{1} \end{array}\right) = \cT_0  \left(\begin{array}{c} \cA_0 \\ \cB_0 \end{array}\right) = \cT_0  \left(\begin{array}{c} 1 \\ \cref \end{array}\right),
\end{equation}
where $\cT_0$ is the transmission matrix from the left half-space into the first layer, $\cref$ is the complex-valued reflection coefficient of the entire layered system, and the incident wave amplitude has been assumed to be unity.  The wave coefficients in layer $MJ$ are related to the coefficients in the right half-space by
\begin{equation}
\left(\begin{array}{c} \cA_{MJ+1} \\ \cB_{MJ+1} \end{array}\right) = \left(\begin{array}{c} \ctran \\ 0 \end{array}\right) = \cT_{MJ}  \left(\begin{array}{c} \cA_{MJ} \\ \cB_{MJ} \end{array}\right),
\end{equation}
where $\cT_{MJ}$ is the transmission matrix from layer $MJ$ into the right half space, $\ctran$ is the complex-valued transmission coefficient.  Relation of the wave across the $MJ+1$ boundaries yields $2 MJ + 2$ linear equations, which is sufficient to solve for the $2 MJ + 2$ unknowns ($\cref$ and $\ctran$, in addition to the $2 MJ$ wave coefficients in the $MJ$ layers).

Upon solving for the unknown quantities, $\cvHl$ in each layer $\ell$ is completely specified and we can succinctly express the total field distribution in the spatial domain as
\begin{equation} \label{Hspacesoln}
\cvH(x,z) = \cH(x,z) \yh = \sum_{\ell=1}^{MJ}
	\rect\left(\frac{z-\zcl}{\dl}\right) \cHl(x,z) \yh,
\end{equation}
where  $\zcl$ is the location of the center of layer $\ell$ and the rect function is defined as
\begin{equation*}
\rect\left(\frac{z-\zcl}{\dl}\right) = \begin{cases}
  1           &  \zcl - \dl/2 \leq z \leq \zcl + \dl/2 \\
  0           &  \mbox{otherwise}.
\end{cases}
\end{equation*}

Equation (\ref{Hspacesoln}) offers a valid representation of the field as a piece-wise function subdivided into spatial intervals corresponding to the layer regions.  Although this form is amenable to numerical routines for solving sets of linear equations, there are at least two disadvantages.  First, compartmentalization of the wave solution into the individual layers does not afford physical insight into the collective behavior of the solution across repeated sets of layers.  Second, representation of the solution in the spatial domain does not produce immediate connections to homogenization parameters, which are generally represented in the spatial-frequency domain.  In the next section, we apply Fourier transformation to the piece-wise wave solution and demonstrate the utility of this strategy for establishing mathematical relations to Floquet-Bloch modes.

\section{Isolating Floquet-Bloch Modes by Fourier Transformation}\label{Fouriersoln}

We re-express the general wave solution given by (\ref{Hspacesoln}) in the spatial-frequency domain by
\begin{equation}\label{FourierGeneral}
\ctH(\kapx,\kapz) = \int_{0}^{\zMJ} \int_{-\infty}^{\infty} \cH(x,z) e^{-i \kapx x} e^{-i\kapz z} dx dz,
\end{equation}
where $\kapx$ and $\kapz$ are the spatial-frequency variables along the respective $x$ and $z$ directions.  Substitution of (\ref{Hspacesoln}) into (\ref{FourierGeneral}) and development of the integrand using well-known Fourier relations and theorems yields
\begin{equation}
\label{Fourier1}
\begin{split}
\ctH(\kapx,\kapz)  =& (2 \pi)^2 \delta(\kapx - \kxo) \sum_{\ell=1}^{MJ} \dl \sinc\left(\frac{\kapz \dl}{2\pi} \right)
	e^{-i\kapz \zcl} \ast \left(\begin{array}{c} e^{-i\ckzl  \zlmo}  \delta(\kapz-\ckzl) \\
		e^{i \ckzl \zlmo} \delta(\kapz + \ckzl) \end{array}\right)^T
	 \left(\begin{array}{c} \cAl \\ \cBl \end{array}\right),
\end{split}
\end{equation}
where $\delta$ is the Dirac delta function.

We next outline a series of mathematical manipulations that can be applied to (\ref{Fourier1}) to produce a physically insightful factorized Fourier-domain wave solution.  First, we re-write the single summation in (\ref{Fourier1}) as a nested double summation over the number of layers in a unit cell and the number of unit cells by making the variable substitutions  $\dl=d_j$ and $\ckzl = \ckzj$ and the index substitution $\ell = mJ + j$, resulting in
\begin{equation}
 \label{Fourier2}
\begin{split}
\ctH(\kapx,\kapz) &= (2 \pi)^2 \delta(\kapx - \kxo) \sum_{m=0}^{M-1}\sum_{j=1}^{J} d_j \sinc\left(\frac{\kapz d_j}{2\pi} \right)
	e^{-i\kapz \zcmj} \ast \\
    &\left(\begin{array}{c} e^{-i\ckzj  \zmjmo}  \delta(\kapz-\ckzj) \\
		e^{i \ckzj \zmjmo} \delta(\kapz + \ckzj) \end{array}\right)^T
	 \left(\begin{array}{c} \cAmj \\ \cBmj \end{array}\right).
\end{split}
\end{equation}
Carrying out the convolution operation in (\ref{Fourier2}) and using the relation $\zcmj-\zmjmo=d_j/2$ yields
\begin{equation}
\label{Fourier3}
\begin{split}
\ctH(\kapx,\kapz) =& (2 \pi)^2 \delta(\kapx - \kxo) \sum_{m=0}^{M-1} \sum_{j=1}^J d_j e^{-i\kapz \zcmj}\\
	& \left(\begin{array}{c} e^{i \ckzj d_j/2} \sinc[(\kapz-\ckzj)d_j/2\pi] \\
		e^{-i\ckzj d_j/2} \sinc[(\kapz+\ckzj)d_j/2\pi]  \end{array}\right)^T
	\left(\begin{array}{c} \cAmj \\ \cBmj \end{array}\right).
\end{split}
\end{equation}The unit cell summation in (\ref{Fourier3}) can be simplified using the relationship
\begin{equation}\label{zcmj}
\zcmj = mD + \zjmo + d_j/2,
\end{equation}
where $D = \sum_{j=1}^J d_j$ is the thickness of the unit cell and $\zjmo$ is the position of the interface between layer $j-1$ and $j$ within unit cell $m=0$.   Substitution of (\ref{zcmj}) into (\ref{Fourier3}) gives
\begin{equation}
\label{Fourier4}
\begin{split}
\ctH(\kapx,\kapz) &= (2 \pi)^2 \delta(\kapx - \kxo) \sum_{j=1}^J d_j e^{-i\kapz \zjmo} \\
	&\left(\begin{array}{c} e^{-i(\kapz-\ckzj) d_j/2} \sinc[(\kapz-\ckzj)d_j/2\pi] \\
		e^{-i(\kapz+\ckzj)d_j/2}  \sinc[(\kapz+\ckzj) d_j/2\pi]  \end{array}\right)^T
	  \sum_{m=0}^{M-1} e^{-i\kapz m D} \left(\begin{array}{c} \cAmj \\ \cBmj \end{array}\right).
\end{split}
\end{equation}

Equation~\ref{Fourier4} expresses the wave solution in terms of wave coefficients $\cAmj$ and $\cBmj$ distributed throughout the medium, which provides little additional insight over the spatial-domain representation of the wave solution in (\ref{Hspacesoln}).  We can further simplify the solution in terms of the wave coefficients in just the first unit cell by using the matrix relationship between wave coefficients in different layers.  The wave coefficients in layer $\ell$ are related to the coefficients in an arbitrary layer $\ell + s$ (where the integer $s \leq MJ - \ell$) within the layered medium by
\begin{equation}
\left(\begin{array}{c} \cAlpN \\ \cBlpN \end{array}\right) =  \cWlNl  \left(\begin{array}{c} \cAl \\ \cBl \end{array}\right),
\end{equation}
where the transfer matrix $\cWlNl$ is determined from the transmission and propagation matrices by
\begin{equation}
\cWlNl =\prod_{q=\ell}^{\ell + s -1} \cTq \cPq.
\end{equation}
The coefficients $\cAmj$ and $\cBmj$ corresponding to layer $j$ within an arbitrary unit cell $m$ can be related to the coefficients $\cAj$ and $\cBj$ corresponding to layer $j$ within unit cell $m=0$ by
\begin{equation}
\left(\begin{array}{c} \cAmj \\ \cBmj \end{array}\right) = \cUj^m  \left(\begin{array}{c} \cAj \\ \cBj \end{array}\right),
\end{equation}
where $\cUj$ is the unit cell transfer matrix from layer $j$ to $j+J$ and can be expressed as
\begin{equation}
\cUj= \cWjJj.
\end{equation}
We can now simplify (\ref{Fourier4}) in terms of wave coefficients $\cAj$ and $\cBj$ distributed throughout a unit cell referenced from layer $j$
\begin{equation}
\label{Fourier5}
\begin{split}
\ctH(\kapx,\kapz) &= (2 \pi)^2 \delta(\kapx - \kxo) \sum_{j=1}^{J} d_j e^{-i\kapz \zjmo}\\
&	\left(\begin{array}{c} e^{-i(\kapz-\ckzj) d_j/2} \sinc[(\kapz-\ckzj) d_j/2\pi] \\
		e^{-i(\kapz+\ckzj) d_j/2} \sinc[(\kapz+\ckzj)d_j/2\pi]  \end{array}\right)^T
	 \left( \sum_{m=0}^{M-1} \left( e^{-i\kapz D} \cUj \right)^m \right) \left(\begin{array}{c} \cAj \\ \cBj \end{array}\right).
\end{split}
\end{equation}

The unit cell transfer matrix referenced from layer $j$, $\cUj$, can be related to the unit cell transfer matrix referenced from layer 1, $\cUo$, using the relation
\begin{equation} \label{Uone}
\cUj^m = \cWjo \,  \cUo^m \,  \cWjo^{-1}.
\end{equation}
Eigenvalue decomposition of $\cUo$ yields
\begin{equation} \label{Uone2}
\cUo=\cQ \; \cl \; \cQ^{-1}
\end{equation}
where $\cQ$ is the eigenvector of $\cUo$ and $\cl$ is the eigenvalue of $\cUo$ whose diagonal elements are the corresponding eigenvalues $\clone$ and $\cltwo$.  Because the determinant of $\cUo$ is unity, the eigenvalues are inverses of each other, $\cltwo=1/\clone$.  As a result, the eigenvalues can be related to the Floquet-Bloch mode, $\ckB$, by
\begin{equation}
\clpm=e^{\pm i \ckB D}.
\end{equation}

To arrive at the final form of the magnetic field solution from (\ref{Fourier4}), we apply eigenvalue decomposition of the unit cell transfer matrix and relate the wave coefficients in layer $j$ to the wave coefficients in the left half space, resulting in
\begin{equation}
\label{Fourier8}
\begin{split}
\ctH(\kapx,\kapz) = & (2 \pi)^2 \delta(\kapx - \kxo)  \sum_{j=1}^{J} \underbrace{ d_j e^{-i\kapz \zjmo} \left(\begin{array}{c} e^{-i(\kapz-\ckzj) d_j/2} \sinc[(\kapz-\ckzj) d_j/2\pi] \\
		e^{-i(\kapz+\ckzj)d_j/2} \sinc[(\kapz+\ckzj)d_j/2\pi]  \end{array}\right)^T \cWjo \cQ }_{\cLj}\\
& \underbrace{\sum_{m=0}^{M-1} \left(e^{-i\kapz D}  \cl \right)^m  }_{\cFB} \underbrace{\cQ^{-1} \cT_0 \left(\begin{array}{c} 1 \\ \cref \end{array}\right)}_{\cC},
\end{split}
\end{equation}
where we have highlighted three distinctive matrix factors - a layer matrix $\cLj$ dependent on the thickness and wave vector in the $j$th layer of the unit cell, a Floquet-Bloch matrix $\cFB$ dependent on the eigenvalues of the unit cell, and a weighting matrix $\cC$ dependent on the reflection coefficient.  Equation~\ref{Fourier8} reveals an inherent relationship between Floquet-Bloch modes and the electromagnetic field in a lossy layered medium of finite extent and is the chief contribution of this work.

The magnetic field solution can be succinctly written as
\begin{equation}\label{compact}
\ctH(\kapx,\kapz) = (2 \pi)^2 \delta(\kapx - \kxo)  \sum_{j=1}^{J} \cLj  \; \cFB \; \cC,
\end{equation}
where the three matrix factors have the following general form
\begin{equation}\label{factors}
\cLj = \begin{pmatrix} \cLone & \cLtwo \end{pmatrix}, \; \cFB = \begin{pmatrix} \cFBone & 0\\ 0 & \cFBtwo\\ \end{pmatrix}, \; \cC =\begin{pmatrix} \cCone \\ \cCtwo \end{pmatrix}.
\end{equation}
In (\ref{factors}), we have distinguished ``forward'' and ``backward'' elements of each matrix factor using the superscript ``+'' and ``$-$'', respectively.  Developing the matrix factors in (\ref{compact}) yields
\begin{equation}\label{expand}
\ctH(\kapx,\kapz) = (2 \pi)^2 \delta(\kapx - \kxo)  \sum_{j=1}^{J} \begin{pmatrix} \cLone \\ \cLtwo\\ \end{pmatrix}^{T}  \; \begin{pmatrix} \cFBone \cCone \\ \cFBtwo \cCtwo\\ \end{pmatrix},
\end{equation}
which, in contrast to the spatial-domain solution in (\ref{Hspacesoln}), requires summation of only $J$ terms corresponding to the layers of the unit cell and describes collective wave behavior across repeated sets of unit cells through the elements of the Floquet-Bloch matrix.

The matrix factors $\cFB$ and $\cC$ appearing in the magnetic field solution can be used to express the corresponding $x$- and $z$-components of the factorized, Fourier-domain electric field solution as
\begin{equation}
\label{Ex}
\begin{split}
\ctEx(\kapx&,\kapz) = (2 \pi)^2 \delta(\kapx - \kxo)  \sum_{j=1}^{J} \frac{1}{\cej \eo \omega} \left(\begin{array}{c} \ckzj \\ -\ckzj \end{array}\right) \cLj \; \cFB \; \cC
\end{split}
\end{equation}
and
\begin{equation}
\label{Ez}
\begin{split}
\ctEz(\kapx&,\kapz) = (2 \pi)^2 \delta(\kapx - \kxo)  \sum_{j=1}^{J} \frac{1}{\cej \eo \omega} \left(\begin{array}{c} \kxo \\ -\kxo \end{array}\right) \cLj \;\cFB\; \cC,
\end{split}
\end{equation}
respectively.

Given the vector spectral magnetic field $\cvH(\kapx,\kapz) = \ctH(\kapx,\kapz) \yh$ and the vector spectral electric field $\cvE(\kapx,\kapz) = \ctEx(\kapx,\kapz) \xh + \ctEz(\kapx,\kapz) \zh$, it is possible to define the spectral time-averaged Poynting vector
\begin{equation}\label{Sspec}
\langle\cvS(\kapx,\kapz)\rangle = \frac{1}{2}\Re\left[\cvE(\kapx,\kapz) \times \cvH^{*}(\kapx,\kapz)\right].
\end{equation}
This is similar to the spectral Poynting vector proposed in Ref.~\cite{Lombardet2005A,Lombardet2005B} to analyze energy propagation of discrete Floquet-Bloch modes in infinite, lossless dielectric photonic crystals, except now extended to accommodate a continuous range of Fourier field components in a finite, lossy periodic system.  It should also be noted that the spectral time-averaged Poynting vector defined in (\ref{Sspec}) is not equivalent to the Fourier transform of the spatial time-averaged Poynting vector, which would involve the convolution of the spectral electric and magnetic fields.  It does, however, enable the spatial frequency $\kapz$ present in the electric and magnetic fields to be envisioned as an electromagnetic plane wave having a well-defined time-averaged direction of power flow.  This concept has recently been applied~\cite{Shakhs2014} to derive band diagrams for layered metamaterials.  Such band diagrams can be used to distinguish forward- and backward-propagating components of the wave in a medium and provide intuitive visualization of phase and power flow.

\section{Limiting Behavior of the Floquet-Bloch Term}

The elements of the Floquet-Bloch matrix are given by
\begin{equation}
\begin{split}
\label{insightful}
\cFBpm =&  e^{-i(\kapz \pm \ckB)  \frac{M-1}{2} D} M \Delta_{\frac{2\pi}{D}}[\kapz \pm \Re(\ckB)] \ast  \sinc\left(\frac{M [\kapz\pm i\Im(\ckB)]D /2}{2\pi}\right),
\end{split}
\end{equation}
where the Dirac comb $\Delta_{\frac{2\pi}{D}}[\kapz\pm\Re(\ckB)]$ is
\begin{equation}
\Delta_{\frac{2\pi}{D}}[\kapz\pm\Re(\ckB)] = \sum_{N=-\infty}^{\infty} \delta\left[\kapz -  2\pi N/D \pm \Re(\ckB)\right],
\end{equation}
describing an infinite comb of discrete spatial-frequency harmonics spaced by $2\pi/D$.  Because the Dirac comb is convolved with the sinc function in (\ref{insightful}), the elements $\cFBpm$ contain harmonics widened through the combined effects of finite extent ($M<\infty$) and material loss ($\Im(\ckB)\neq0$).  The forward element $\cFBone$ has a principal harmonic centered at $\Re(\ckB)$ and the backward element $\cFBtwo$ has a principal harmonic centered at $-\Re(\ckB)$.

For a medium of infinite extent ($M \rightarrow \infty$), the elements of the Floquet-Bloch matrix approach
\begin{equation}
\begin{split}
&\lim_{M \rightarrow \infty}  \cFBpm =
\begin{cases}
  M \Delta_{\frac{2\pi}{D}}(\kapz \pm \ckB) &  \Im(\ckB)=0\\
  \Delta_{\frac{2\pi}{D}}[\kapz \pm \Re(\ckB)] \ast \frac{2}{D \sqrt{\kapz^2 + \Im(\ckB)^2}} & \Im(\ckB)\neq 0.
\end{cases}
\end{split}
\end{equation}
In the absence of loss, $\cFBpm$ are discrete spectra with peaks at harmonics of $\pm\Re(\ckB)$.  The resulting magnetic field solution consists of discrete forward and backward Floquet-Bloch modes (equivalent to the classical Floquet-Bloch solution) with amplitudes that can be explicitly determined by $(2 \pi)^2 \sum_{j=1}^{J} \cLone \cCone$ and $(2 \pi)^2 \sum_{j=1}^{J} \cLtwo \cCtwo$, respectively.  In the presence of loss, $\cFBpm$ are continuous spectra with peaks that are centered about harmonics of $\pm\Re(\ckB)$ and broadened due to the effect of $\Im(\ckB)$.

\section{Layered Metamaterial Analysis}  \label{AnalysisSec}

\begin{figure*}
\begin{center}
\includegraphics[scale=0.80]{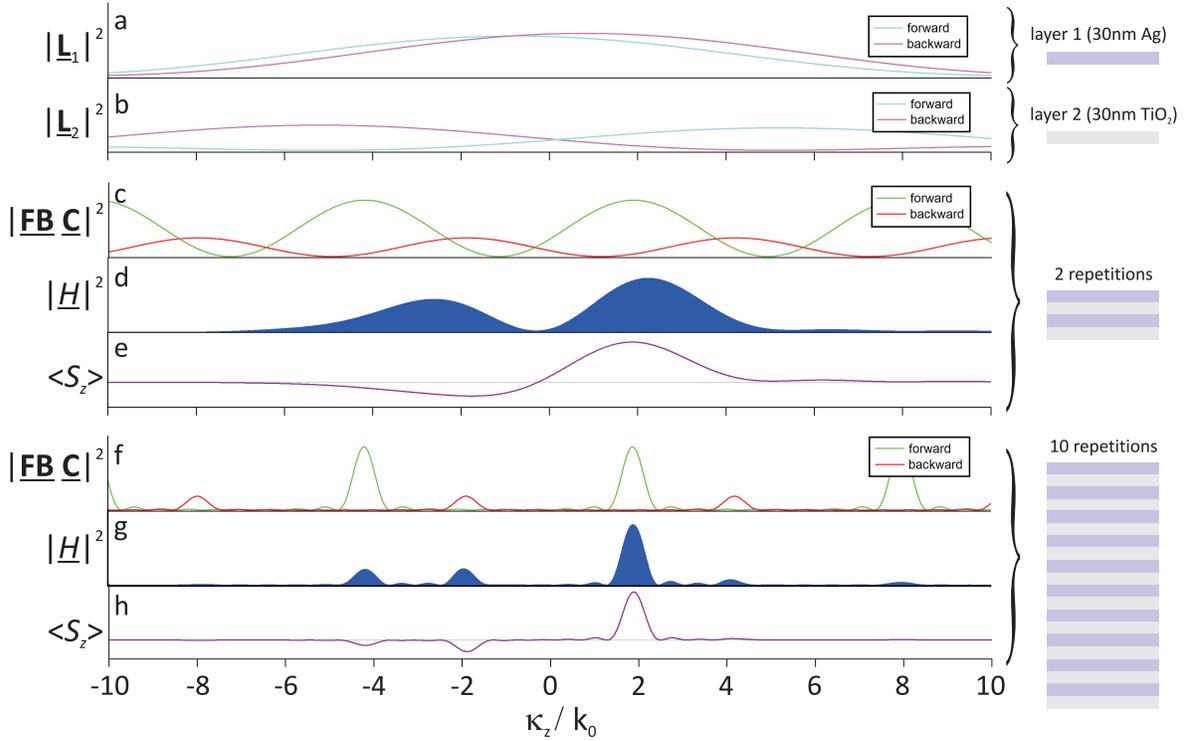}
\caption{\label{MetaSoln} Decomposition of the wave solution in a metal-dielectric bi-layer system consisting of alternating layers of  $30$-$\nm$-thick Ag and  $30$-$\nm$-thick TiO$_2$, assuming a normally incident TM-polarized wave with a free-space wavelength of $\lo = 365\nm$.  a) The forward and backward components of the layer matrix $|\cLMat_{1}|^2$ corresponding to the $30$-$\nm$-thick Ag layer.  b)  The forward and backward components of the layer matrix $|\cLMat_{2}|^2$ corresponding to the $30$-$\nm$-thick TiO$_2$ layer. c), d), and e) depict the forward and backward components of $|\cFB \, \cC|^2$, the magnetic field spectrum $|\ctH|^2$, and the $z$-component of the time-averaged spectral Poynting vector, respectively, for the case of 2 unit cells;  f), g), and h) depict the same set of information for the case of 10 unit cells.   The horizontal gray lines in e) and h) correspond to zero values of the spectral Poynting vector.}
\end{center}
\end{figure*}

\begin{figure}
\begin{center}
\includegraphics[scale=0.35]{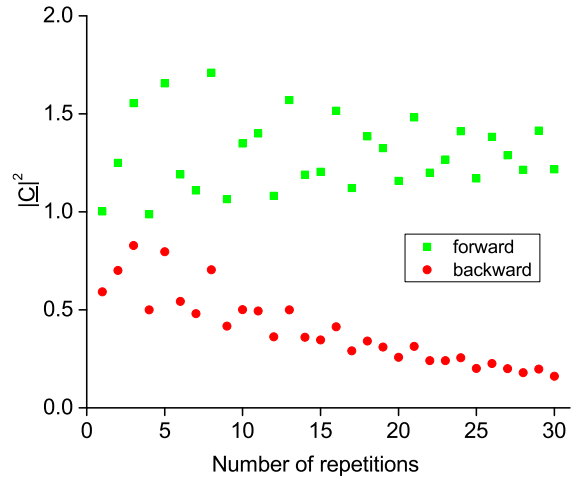}
\caption{\label{MetaCoeff}Forward and backward components of the weighting matrix $|\cC|^2$ versus the number of repetitions.  Here, we have assumed a metal-dielectric bi-layer system with a unit-cell made from a $30$-$\nm$-thick Ag layer and $30$-$\nm$-thick TiO$_2$ layer, assuming a normally incident TM-polarized wave with a free-space wavelength of $\lo = 365\nm$.}
\end{center}
\end{figure}

We consider a typical layered metamaterial structure consisting of metal-dielectric layers subject to normal-incidence, TM-polarized illumination ($\kxo = 0$).  If the system consists of a bi-layer unit cell, the factorized, Fourier-domain magnetic field solution takes the form
\begin{equation}\label{bilayereqn}
\ctH(\kapz) = 2 \pi \begin{pmatrix} \cL^+_1 + \cL^+_2 \\ \cL^-_1 + \cL^-_2\\ \end{pmatrix}^{T}  \; \begin{pmatrix} \cFBone \cCone \\ \cFBtwo \cCtwo\\ \end{pmatrix},
\end{equation}
where $\cL^\pm_{1}$ and $\cL^\pm_{2}$ are the forward and backward elements of layer matrices respectively associated with layers 1 and 2 of the unit cell.  Let's take the case bi-layer unit cell comprising a $30$-$\nm$-thick Ag layer and a $30$-$\nm$-thick TiO$_2$ layer.  We will consider a system composed of two unit cells ($M=2$) and another composed of ten unit cells ($M=10$).  The systems are bounded by free space and excited by a  wave with a free-space wavelength $\lo = 365\nm$.  We assume that Ag has a complex refractive index of $0.076 + 1.605 i$ (interpolated from experimental data~\cite{J1972}) and that TiO$_2$ has a real refractive index of 2.80.

Figure \ref{MetaSoln} highlights, for the $M=2$ and $M=10$ cases, the modulus squared of the magnetic field solution and the elements of its matrix factors, in addition to the $z$-component of the time-averaged spectral Poynting vector.  The magnetic field spectrum results from the confluence of broad spectral envelopes defined by $\cL^\pm_{1}$ (Ag layer) and $\cL^\pm_{2}$ (TiO$_2$ layer) and finer spectral combs defined by $\cFBone \cCone$ and $\cFBtwo \cCtwo$.  The Floquet-Bloch mode of the unit cell is $\ckB = 32.5 + 0.4i \um^{-1}$, corresponding to a  Floquet-Bloch refractive index $\cnB = 1.89 + 0.02 i$.  The combs defined by $\cFBone \cCone$ and $\cFBtwo \cCtwo$ are offset; the former with a principal peak located at $\Re(\ckB)$ and the latter with a principal peak located at $-\Re(\ckB)$.  Increasing $M$ from 2 to 10 narrows the peaks of the combs, which in turn narrows the peaks in the magnetic field spectrum.  Due to the losses in the metallic layer, the spectral peaks defined by $\cFBone \cCone$ and $\cFBtwo \cCtwo$ have finite width even in the limit of an infinite system, underscoring the limitation of the classical Floquet-Bloch solution for describing the spectral composition of the wave in lossy periodic systems.

For both the $M=2$ and $M=10$ cases, the magnitude of $\cFBone \cCone$ is larger than that of $\cFBtwo \cCtwo$, meaning that the forward propagating components in the medium outweigh the backward propagating components.  As shown in Figure~\ref{MetaCoeff}, $|\cCone|^2$ generally exceeds $|\cCtwo|^2$, with the latter gradually approaching zero as the number of repetitions increases.  It is noteworthy that the negative spatial-frequency components of the wave solution are characterized by a time-averaged spectral Poynting vector that is negative, suggesting that these wave components are actually forward-propagating waves (having parallel phase and energy velocities) that have been reflected in the system.  This is counter to the widely held belief that the negative spatial-frequency components of a Floquet-Bloch field decomposition are backward-propagating waves~\cite{Ramo1965}.

The Floquet-Bloch refractive index can take on values with negative real parts, particularly for TM-polarized illumination of finely layered structures composed of metal at frequencies near the bulk plasma frequency of the metal~\cite{Lezec2013,V2010}.  Such layered media are capable of mimicking refractive properties associated with negative-index media such as flat-lens imaging with a planar slab~\cite{Lezec2013,Ott2014}.  The physical significance of the sign of the real part of the Floquet-Bloch refractive index, however, is not immediately clear.  As seen here, the magnetic field spectrum is generally dispersed in peaks spanning positive and negative spatial frequencies and a number of factors conspire to determine the relative weights of these peaks.  Moreover, in systems composed of right-handed materials, it has been consistently observed that the positive and negative spatial-frequency components of the wave are both forward-propagating waves.

\section{Conclusion}

We have developed a new Fourier-domain representation for the electromagnetic field solution in a generic layered medium under the constraints of loss and finite extent.  The main contribution is the identification and isolation of a term explicitly dependent on the Floquet-Bloch mode using only Maxwell's equation and without a priori invoking the Floquet-Bloch theorem.  The Floquet-Bloch modes in the medium are complex-valued and generally exist as a continuum whose density is determined by boundary reflections and layer composition.  We have condensed the Fourier-domain field solution into a compact product of three terms, each dependent on distinctive physical parameters of the system.  This work demonstrates the utility of Fourier-domain analysis of electromagnetic field solutions, which will be extended to more complex geometries in future studies.

\section*{Acknowledgments}
We acknowledge helpful discussions with Prof. Thomas Johnson from UBC.  This work was supported by Natural Sciences and Engineering Research Council of Canada (NSERC) Discovery Grant 366136.

\end{document}